\documentclass[iicol,pdflatex,sn-nature]{sn-jnl}% Style for submissions to Nature Portfolio journals
\usepackage{graphicx}%
\usepackage{multirow}%
\usepackage{amsmath,amssymb,amsfonts}%
\usepackage{amsthm}%
\usepackage{mathrsfs}%
\usepackage[title]{appendix}%
\usepackage{xcolor}%
\usepackage{textcomp}%
\usepackage{manyfoot}%
\usepackage{booktabs}%
\usepackage{algorithm}%
\usepackage{algorithmicx}%
\usepackage{algpseudocode}%
\usepackage{listings}%
\theoremstyle{thmstyleone}%
\theoremstyle{thmstyletwo}%

\theoremstyle{thmstylethree}%

\begin{document}

\title[Turning AI Data Centers into Grid-Interactive Assets]{Turning AI Data Centers into Grid-Interactive Assets: Results from a Field Demonstration in Phoenix, Arizona}


\author[1]{\fnm{Philip} \sur{Colangelo}}%\email{philip.colangelo@emeraldai.co}
\author*[1]{\fnm{Ayse K.} \sur{Coskun}}\email{ayse.coskun@emeraldai.co}
\author[1]{\fnm{Jack} \sur{Megrue}}%\email{jack.megrue@emeraldai.co}
\author[1]{\fnm{Ciaran} \sur{Roberts}}%\email{ciaran.roberts@emeraldai.co}
\author[1]{\fnm{Shayan} \sur{Sengupta}}%\email{shayan.sengupta@emeraldai.co}
\author[1]{\fnm{Varun} \sur{Sivaram}}%\email{varun@emeraldai.co}
\author[1]{\fnm{Ethan} \sur{Tiao}}%\email{ethan.tiao@emeraldai.co}
\author[1]{\fnm{Aroon} \sur{Vijaykar}}%\email{aroon.vijaykar@emeraldai.co}
\author[1]{\fnm{Chris} \sur{Williams}}%\email{chris.williams@emeraldai.co}
\author[1]{\fnm{Daniel C.} \sur{Wilson}}%\email{daniel.wilson@emeraldai.co}

\author[2]{\fnm{Zack} \sur{MacFarland}}%\email{zmacfarland@nvidia.com}

\author[3]{\fnm{Daniel} \sur{Dreiling}}\nomail
\author[3]{\fnm{Nathan} \sur{Morey}}%\email{Nathan.Morey@srpnet.com}

\author[4]{\fnm{Anuja} \sur{Ratnayake}}
\author[4]{\fnm{Baskar} \sur{Vairamohan}}

\affil*[1]{\orgname{Emerald AI, authors are listed in alphabetical order}}
\affil*[2]{\orgname{NVIDIA Corporation}}
\affil*[3]{\orgname{Salt River Project (SRP)}}
\affil*[4]{\orgname{Electric Power Research Institute (EPRI)}}


\abstract{Artificial intelligence (AI) is fueling exponential electricity
demand growth, threatening grid reliability, raising prices for communities
paying for new energy infrastructure, and stunting AI innovation as data
centers wait for interconnection to constrained grids. This paper presents the
first field demonstration, in collaboration with major corporate partners, of a
software-only approach--Emerald Conductor--that transforms AI data centers into
flexible grid resources that can efficiently and immediately harness existing
power systems without massive infrastructure buildout. Conducted at a 256-GPU
cluster running representative AI workloads within a commercial, hyperscale
cloud data center in Phoenix, Arizona, the trial achieved a 25\% reduction in
cluster power usage for three hours during peak grid events while maintaining
AI quality of service (QoS) guarantees. By orchestrating AI workloads based on
real-time grid signals without hardware modifications or energy storage, this
platform reimagines data centers as grid-interactive assets that enhance grid
reliability, advance affordability, and accelerate AI's development.}




\maketitle

\clearpage

\section{Introduction}\label{sec:intro}
The global proliferation of AI technologies has led to surging demand for high-performance data centers, increasingly powered by GPU clusters~\cite{johnson2025implications}. As these AI clusters scale, their energy consumption poses a growing strain on power grids~\cite{bianchini2024data}--particularly during periods of high demand or low renewable output. In the United States alone, projections estimate that AI-related data center demand could reach tens of gigawatts by 2030, exacerbating grid congestion and delaying project deployments~\cite{iea2025electricity,aljbour2024poweringintelligence}.

Historically, demand response in data centers has been explored in academic settings, mostly using CPU-based clusters running high performance computing (HPC) applications~\cite{zhang2022aqa} or demonstrating the potential of demand response via simulation and analytical models~\cite{zhang2021realworld,xing2023carbonresponder}. These studies provided valuable insights but did not account for the rigid performance demands and distinct energy profiles of AI training and inference workloads on GPUs. Other work has demonstrated that data centers can reduce operational carbon emissions by allocating fewer compute resources to jobs broadly classified as having flexible performance needs~\cite{radovanovic2023carbon}.

Scalable, software-based solutions that respect AI service-level agreements (SLAs) while offering real-time power modulation are urgently needed. Our central hypothesis is that GPU-driven AI workloads contain enough operational flexibility--when smartly orchestrated--to participate in demand response and grid stabilization programs~\cite{sivaram2018taming}.
Although utilities offer financial incentives for power flexibility, other adoption costs--such as impacts on workload performance and delays in deploying new data centers--can limit participation~\cite{tansoo2024usingcba}, especially amid rapid AI growth.
Utilities and system operators can further prioritize flexible AI data centers for accelerated interconnection and offer them lower tariffs and flexibility payments, recognizing their benefits to system reliability and ability to utilize existing system headroom, estimated at 100GW in the United States~\cite{norris2025rethinking}.

In this paper, we present results from the first real-world demonstration of a software-based platform, Emerald Conductor, which transforms a production AI cluster into a grid-responsive asset. Conducted in Phoenix, Arizona by Emerald AI in partnership with Oracle Cloud Infrastructure, NVIDIA, and regional utility Salt River Project (SRP) through the EPRI DCFlex initiative, this test accomplishes the first validation of providing sustained, accurate power reductions using only workload orchestration--without requiring any hardware retrofits or energy storage systems.

\section{Results}\label{sec:results}

\subsection{System Architecture and Flexibility Framework}
\begin{figure}[t]
	\centering
	\includegraphics[width=.9\linewidth]{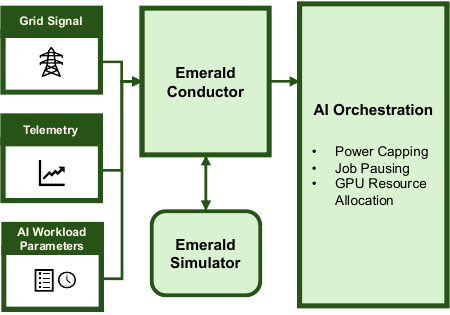}
	\caption{Overview of the Emerald Conductor architecture.}\label{fig:overview}
\end{figure}

At the core of the demonstration is \textbf{Emerald Conductor}, a software
platform that interfaces with AI workload managers and grid signal sources.
Conductor dynamically schedules jobs, modifies resource allocations for each
job, and applies power-limiting techniques such as GPU frequency scaling. To
guide its decisions, Conductor uses the \textbf{Emerald Simulator}, a
system-level model trained to predict the power-performance behavior of AI
jobs. The simulator evaluates the trade-offs of various orchestration
strategies under operational constraints and grid needs, recommending an
orchestration strategy to assure AI workload QoS guarantees while also meeting
power grid response commitments.  Fig.~\ref{fig:overview} demonstrates the
overall architecture of Emerald Conductor and Emerald Simulator.

Our \textbf{workload tagging schema} classifies jobs into flexibility tiers based on
user tolerance for runtime or throughput deviations. The tags allow applying
control policies that differentiate jobs and ensure each job meets the compute
user's desired QoS threshold, which could pave the way for an SLA that upholds
user-specified QoS while enabling flexible power management. For example, using
feedback and guidance from our industry partners, we identified the following
flexible SLAs for representative AI workloads: (a) \textbf{Flex 0}: no performance
reduction (strict SLA); (b) \textbf{Flex 1}: up to 10\% performance (average throughput)
reduction allowed over a 3-6 hour period; (c) \textbf{Flex 2}: up to 25\% allowed; (d)
\textbf{Flex 3}: up to 50\% allowed. These tiers enable intelligent, non-disruptive
throttling that preserves workload commitments while unlocking power
flexibility.

\subsection{Phoenix Field Trial}
The demonstration was conducted at an \textbf{Oracle} Phoenix Region Cloud data
center on a 256-GPU cluster built on \textbf{NVIDIA A100} Tensor Core GPUs,
orchestrated through \textbf{Databricks MosaicML}
(\url{https://www.databricks.com/research/mosaic}), instrumented via \textbf{Weights
\& Biases} (\url{https://wandb.ai/}) for telemetry, and integrating
\textbf{Amperon's} grid demand forecasting tools. Four representative workload
ensembles were selected--each combining varying proportions of training,
inference, and fine-tuning jobs (see \nameref{sec:methods} for more details).

\begin{figure*}[t]
	\centering
	\includegraphics[width=.95\textwidth]{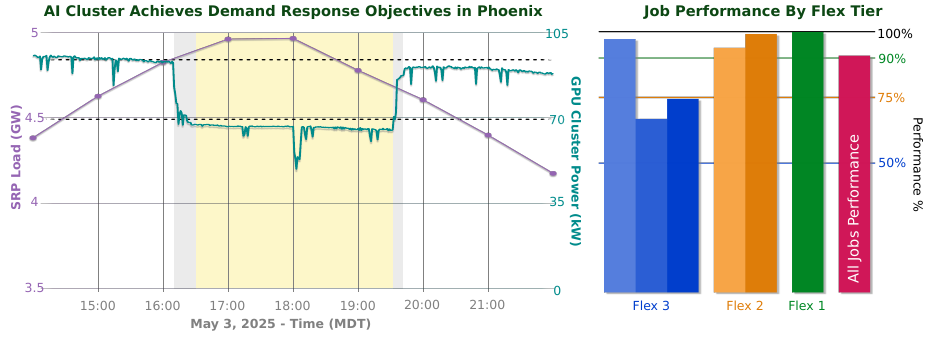}
	\caption{SRP event summary figure shows the power reduction curve and
	applications with different flexible SLAs meeting their performance
	requirements. The power reduction was achieved using a mixture of control knobs
	and power capping (DVFS + Job Pausing, Fair--see
	\nameref{ssec:algorithms}).}\label{fig:srpsummary}
\end{figure*}

\begin{figure*}[t]
\centering
\includegraphics[width=.95\textwidth]{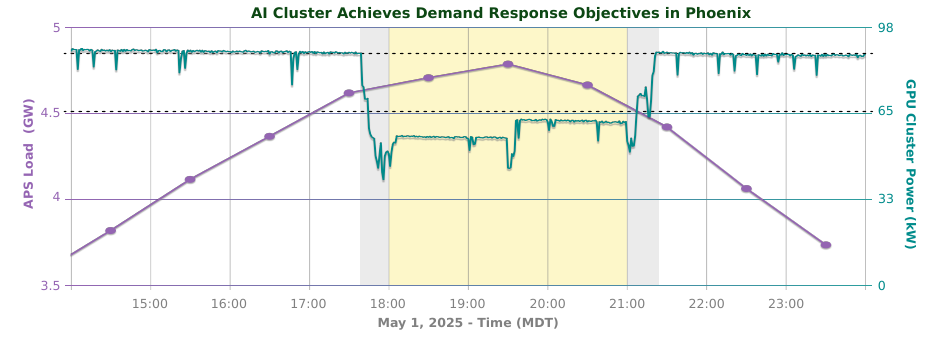}
\caption{APS event showing power reduction curve achieved by Emerald Conductor. The power reduction was achieved using a mixture of control knobs without power capping (Job Pausing + Resource Allocation, Fair--see \nameref{ssec:algorithms}).}\label{fig:apssummary}
\end{figure*}

In consultation with the regional utilities Arizona Public Service (APS) and
Salt River Project (SRP), we set stringent targets for grid-responsive demand
to prove that AI compute power load could provide meaningful relief during
periods of system-coincident peak stress, for example during a hot Phoenix day
with high air conditioning load. We executed two events to demonstrate these
capabilities: (a) \textbf{May 1, 2025 (addressing APS system peak)}, and (b)
\textbf{May 3, 2025 (addressing SRP system peak)}, where we tracked the grid
load and identified the upcoming peak demand periods. Each event required the
cluster to reduce power by \textbf{25\%} with respect to the average base load
during the peak demand period, sustain the reduction for \textbf{3 hours}, and
\textbf{ramp down and up gracefully over 15 minutes}, avoiding so-called ``snap
back'' at the conclusion of the event by staying below the pre-event baseline.
These and other technical requirements of the test were set by our utility
partners. On both occasions, Emerald Conductor met the utility-set requirements
precisely (Figs.~\ref{fig:srpsummary} and~\ref{fig:apssummary}).

In both APS and SRP events, all jobs completed within allowable SLA envelopes.
We confirmed via power measurements that our software solution achieved
sustained and accurate reductions.

\begin{figure*}[t]
	\centering
	\includegraphics[width=.95\textwidth]{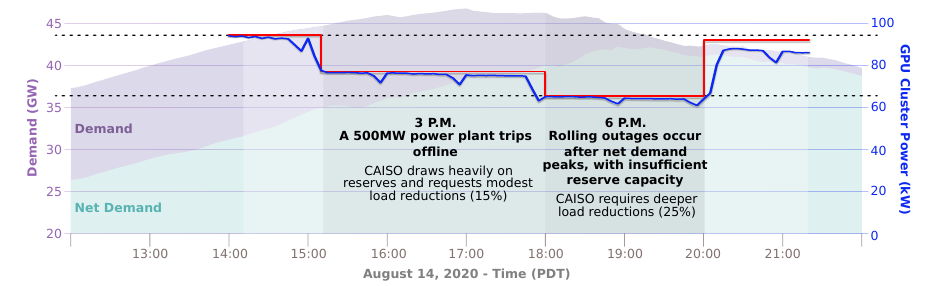}
	\caption{Reenacted CAISO power event and how Emerald Conductor responded to the curtailment requirement. Power consumption shown in the plot is measured and averaged at continuous 5-minute windows.}\label{fig:caisoevent}
\end{figure*}

We also modeled a synthetic emergency event as part of the field trial, based
on California ISO (CAISO)'s August 2020 emergency load shed event, where a
power plant failure during an extreme weather event triggered CAISO to request
reduction of power via available reserves in the grid~\cite{caiso2021rca}. In this reenactment,
Conductor responded to an initial 15\% curtailment followed by an emergency 10\%
further step-down. The system delivered both reductions smoothly, matching the
desired power profile (Fig.~\ref{fig:caisoevent}), while continuing to assure the AI QoS
thresholds.

In addition to demonstrating capabilities for APS, SRP, and CAISO sample
events, we ran a total of 33 experiments (each 3-6 hours long) including 212
total individual jobs during the field trial, where we demonstrated the impact
of applying several different power management policies on our select workload
ensembles (see \nameref{sec:methods} for further details). In every single experiment,
Emerald Conductor performed as expected, guided by the Emerald Simulator
predictions to achieve the required power reduction and job-specific SLA
requirements.

\subsection{Simulator Performance Accuracy}
\begin{figure*}[t]
	\centering
	\includegraphics[width=.95\textwidth]{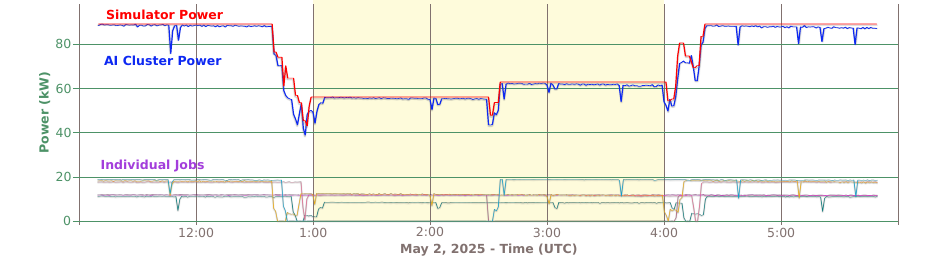}
	\caption{Power prediction of Emerald Simulator vs. measured power during a power reduction event. The bottom part of the plot also shows the individual controls applied to each job in the workload ensemble running during this event.}\label{fig:simvsmeasured}
\end{figure*}

The Emerald Simulator's predictions closely matched real-world cluster
behavior. Across control intervals in our experiments, the model achieved
4.52\% root-mean-square error (RMSE) in power predictions, relative to average
experiment power (Fig.~\ref{fig:simvsmeasured}). Individual job behaviors, including fine-tuning and
batch inference workloads, were predicted with sufficient fidelity to inform
real-time orchestration without breaching SLAs.

\section{Discussion}\label{sec:discussion}
\subsection{Data Centers as Grid Assets}
This demonstration marks a paradigm shift in the role of AI data centers--from
static, high-load consumers to \textbf{active, controllable grid participants}. By
integrating software intelligence with workload management, AI clusters can
shape their power consumption profiles dynamically in response to grid needs.

Recent studies demonstrate that load flexibility for AI data centers to reduce
power use by roughly 25\% for up to 200 hours a year, or far less than 1\% of
the time, could unlock \textbf{up to 100 GW of new data center capacity} in the
U.S.  without requiring extensive new generation or transmission
infrastructure~\cite{norris2025rethinking,lin2024explodingai}--enough to meet projected AI
growth for the next decade.

\subsection{Scalability and Deployment}
A central advantage of Emerald Conductor is \textbf{deployability on standard
hardware}. Unlike hardware retrofits or battery deployments, which incur
significant capital costs and operational complexity, this software platform
runs atop existing cloud and HPC stack environments. This means the approach is
\textbf{cloud-native}, scalable, and already compatible with emerging AI
cluster standards, including NVIDIA's modern infrastructure and containerized
ML platforms. An opportunity exists to deploy this method in other major AI
data center regions, particularly those with constrained grid conditions like
Northern Virginia, Silicon Valley, and
Texas~\cite{aljbour2024poweringintelligence}, and globally in countries with
strong data center growth such as the UK, Ireland, Germany, Singapore, and
others~\cite{iea2025electricity}.

\subsection{Control Knob Tradeoffs}
\begin{figure}[t]
	\centering
	\includegraphics[width=\linewidth]{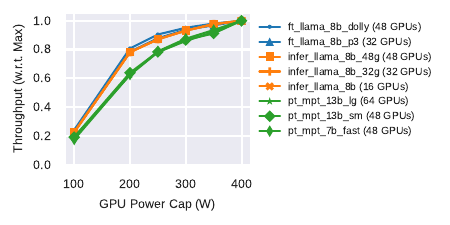}
	\caption{Throughput of power-capped AI workloads on the 256 GPU cluster.}\label{fig:dvfs}
\end{figure}

Our demonstration explored several \textbf{control knobs} to regulate power
consumption: (a) \textbf{power capping} via the NVIDIA-System Management
Interface (SMI), which primarily uses dynamic voltage frequency scaling
(\textbf{DVFS}) to effectively reduce power with minor throughput impacts for
many workloads~\cite{zhao2023sustainable} (Fig.~\ref{fig:dvfs}); (b) \textbf{job pausing
(de-prioritization)}, which temporarily pauses running jobs for steep
reductions in power, and (c) \textbf{changing the allocated resources for
jobs}, which reduces the number of allocated GPUs to reduce power while
allowing job progress.

We evaluated performance sensitivity to power capping across multiple GPU
allocations and workload types, including: fine-tuning (ft) LLaMA-8B on two
datasets, LLaMA-8B inference (infer) across three different GPU allocation
sizes, and three configurations of Mosaic Pretrained Transformer (MPT)
pre-training (pt) workloads (see \nameref{ssec:workloads} section for details).
Across all experiments, job performance exhibited sensitivity to power limits,
with modest degradation observed as power caps approached 400 W--the typical
thermal design power (TDP) of the target GPU. As power caps decreased further
below this threshold, performance degradation became more pronounced. The
number of GPUs allocated had minimal influence on power sensitivity; however,
power-performance tradeoffs varied by workload. In particular, MPT pre-training
jobs were notably more sensitive in the mid-range of power caps, showing
greater performance drops compared to fine-tuning and inference tasks.

Control overhead of power capping is negligible, making real-time
responsiveness feasible even in busy clusters. Both pausing and re-allocating
resources for running jobs require checkpointing for training jobs to ensure
forward progress and minimize the overhead associated with these knobs.
Although prior work offers methods to optimize for long-term objectives where
changes to control state incur a cost~\cite{lechowicz2023pauseresume}, we are
able to treat checkpointing overhead as negligible for our relatively
infrequent demand response events since training jobs often run for days (or
even weeks) at a time. An essential ingredient in Emerald Conductor is to apply
these knobs in a carefully calculated manner to avoid SLA
violations--therefore, our software was able to meet SLAs even considering the
more conservative experimental duration of only 6 hours.

In experiments, our ``DVFS + Job Pausing, Fair'' policy, which spreads
performance slowdowns across all jobs proportionally with their flexible SLAs
(see \nameref{sec:methods} for more details), provided a good balance between
achieving higher average job throughput and reducing the number of jobs
impacted by power management.

\subsection{Policy and Market Implications}
Software-defined flexibility in AI data centers could transform \textbf{utility
interconnection processes}. Flexibility may accelerate data center
interconnection and siting approvals, particularly in regions where permitting
delays have become a
bottleneck~\cite{ratnayake2025gridflexibilityneeds,satchwell2025electricity}.
Data centers that are able to respond to grid signals can also avoid expensive
infrastructure upgrades and receive \textbf{demand response credits} in
capacity or ancillary markets.  Moreover, flexible data centers could
complement renewable energy integration by reducing demand during periods of
low intermittent renewable production and matching ramping requirements of
fossil backup plants.

\subsection{Limitations}
Several limitations remain in the current approach to grid-interactive AI data
centers. First, not all AI workloads are temporally flexible; AI customers with
jobs with strict latency or reliability requirements may not be willing to
accept workload throttling at a single site. In this study, batch-style
training, fine-tuning, and inference tasks could be slowed or paused, while
other ``Flex 0'' jobs such as real-time inference, streaming, and model serving
were not modified. To unlock broader flexibility, future work will explore
geographically shifting AI
workloads~\cite{zheng2020mitigatingcurtailment,mehra2023supporting}, harnessing
spatiotemporal flexibility to preserve performance with minimal latency penalty
across the broadest range of AI workloads. In addition, the industry's
incentives and SLAs will need to evolve, encouraging users to opt into
flexibility tiers in exchange for cost or compute availability benefits that
are enabled by the massive new capacities of AI compute that can be brought
online to power grids thanks to flexibility.

Our field demonstration focused on a single cluster's ability to curtail load
to respond to system peaks and reduce the system--coincident peak demand of an
AI data center, enabling faster interconnection and avoiding infrastructure
buildout to serve higher system peaks. To fully understand system-level
impacts, larger-scale deployments involving full data center telemetry and
multiple data center zones are needed. These would allow for the validation of
broader control strategies and coordination mechanisms. Our future work will
also explore participation, in addition to emergency demand response, in a
wider range of grid programs such as day-ahead demand response, frequency
regulation, and other ancillary
services~\cite{zhou2016survey,ercot2024ancillary} to assess economic viability
and technical readiness in real-world grid markets.

\section{Methods}\label{sec:methods}
\subsection{Workloads}\label{ssec:workloads}
For our field trial, we worked with Databricks to design four representative
workload ensembles consisting of training, inference, and fine-tuning jobs. We
downloaded MPT and LLaMA 3.1 Family models from Hugging Face and used the C4
dataset for pretraining and Dolly and P3 datasets for fine-tuning. These
workloads and their flexible SLA tiers are shown in Table~\ref{tab:ensembles}.
Each job was tagged into one of four flexibility tiers (discussed earlier in
Results), based on tolerance to runtime or throughput degradation. We
determined these ensembles and tags in consultation with our industry expert
partners and based on typical SLA expectations for different types of AI jobs.

\begin{table*}[t]
\caption{Workload ensembles and their flexibility SLAs used in the experiments.
	In the cluster under test, each GPU node consisted of 8 A100 GPUs, and
	each workload ensemble ran throughout the experiment duration, with
	control actions determined by Emerald Conductor.}\label{tab:ensembles}
\centering
\begin{tabular}{llcl}
\toprule
\textbf{Ensemble} & \textbf{Workload} & \textbf{\# Nodes} & \textbf{Flex Level} \\
\midrule

\multirow{6}{*}{\shortstack[l]{Ensemble 1 \\ 80\% training,~20\% inference}}
  & MPT 13B - training     & 8 & Flex 3 \\
  & MPT 7B - training      & 6 & Flex 3 \\
  & LLaMA 8B - finetune    & 6 & Flex 2 \\
  & LLaMA 8B - finetune    & 6 & Flex 3 \\
  & LLaMA 8B - inference   & 4 & Flex 0 \\
  & LLaMA 8B - inference   & 2 & Flex 0 \\

\midrule

\multirow{6}{*}{\shortstack[l]{Ensemble 2 \\ 50\% training,~50\% inference}}
  & MPT 13B - training     & 8 & Flex 3 \\
  & MPT 7B - training      & 6 & Flex 3 \\
  & LLaMA 8B - finetune    & 4 & Flex 2 \\
  & LLaMA 8B - inference   & 4 & Flex 0 \\
  & LLaMA 8B - inference   & 6 & Flex 0 \\
  & LLaMA 8B - inference   & 4 & Flex 0 \\

\midrule

\multirow{6}{*}{\shortstack[l]{Ensemble 3 \\ 50\% training,~50\% inference}}
  & MPT 13B - training     & 6 & Flex 3 \\
  & MPT 7B - training      & 6 & Flex 3 \\
  & LLaMA 8B - finetune    & 4 & Flex 2 \\
  & LLaMA 8B - inference   & 4 & Flex 3 \\
  & LLaMA 8B - inference   & 6 & Flex 2 \\
  & LLaMA 8B - inference   & 6 & Flex 1 \\

\midrule

\multirow{8}{*}{\shortstack[l]{Ensemble 4 \\ 90\% training,~10\% inference}}
  & MPT 13B - training     & 6 & Flex 3 \\
  & MPT 7B - training      & 6 & Flex 3 \\
  & LLaMA 8B - finetune    & 4 & Flex 2 \\
  & LLaMA 8B - finetune    & 4 & Flex 3 \\
  & LLaMA 8B - inference   & 2 & Flex 0 \\
  & LLaMA 8B - inference   & 2 & Flex 0 \\
  & LLaMA 8B - finetune    & 4 & Flex 3 \\
  & LLaMA 8B - finetune    & 4 & Flex 2 \\

\bottomrule
\end{tabular}
\end{table*}

\subsection{Orchestration Algorithms}\label{ssec:algorithms}
We designed a suite of power and load management algorithms for Emerald
Conductor. These algorithms utilize the three control knobs described
earlier--both individually and in various combinations (e.g., DVFS with job
pausing, job pausing with resource reallocation, or others).

We implemented two main algorithmic strategies across different knob
combinations: \textbf{(a) Greedy}: This algorithm prioritizes applying control
knobs to the most flexible jobs first, aiming to maximize power reduction while
minimizing the number of jobs affected. \textbf{(b) Fair}: This algorithm
distributes the projected performance overhead more evenly across all jobs,
ensuring a more balanced impact on workload performance.

All policies we implemented solved for \textbf{power reduction objectives},
subject to constraints from job SLAs. The following ``control knob, algorithmic
strategy'' combinations provided the most desirable results while meeting power
reduction and job performance constraints:

\begin{itemize}
	\item \textit{DVFS + Job Pausing}, Fair provided the best tradeoff between average job throughput and the number of jobs impacted;
	\item \textit{DVFS, Fair} provided the best average job throughput overall as the knob can be applied at a finer granularity than other knobs;
	\item \textit{Job Pausing, Greedy} affected the fewest number of jobs to achieve the desired power reduction;
	\item \textit{Job Pausing + Resource Allocation, Fair} achieved the best average throughput among policies without DVFS.
\end{itemize}

\subsection{Simulation and Real-Time Execution}
In this field trial, we profiled each job in advance of the power reduction
event experiments, and used estimated power-performance relationships based on
these profiles to determine the policy decisions in the Emerald Simulator. The
Simulator then drove Conductor's decisions at runtime. We received the grid
load signals via API from the Amperon platform, which provides forecast grid
load and actual historical grid data that we used to verify the timing of our
cluster's demand response performance. We timestamp-aligned all power and job
completion telemetry.

\subsection{Evaluation Metrics}
Three key metrics were used to assess performance in our experiments (a): \textbf{power
reduction compliance}, which compares percentage power reduction achieved vs.
utility target; (b) \textbf{QoS preservation}, which checks whether individual job SLAs
are met, and (c) \textbf{simulator accuracy}, where we calculate the root mean square
error (RMSE) of power prediction vs measured power. We measured power
consumption of GPUs via NVDIA-SMI and throughput (e.g., steps per second or
tokens per second) of applications via our custom scripts.

In all power reduction events (APS, SRP, CAISO) and our 33 total power
reduction experiments, we fully maintained compliance thresholds while
maintaining zero SLA violations. Across our experiments, we achieved 4.52\%
simulator accuracy RMSE relative to average experiment power.

\section{Implementation of Emerald Conductor}
We implement the Emerald Conductor software as a centralized Python application
that issues commands to compute-node management processes hosted alongside
executing jobs. Conductor starts, stops, and queries the scheduling state of
jobs through the MosaicML mcli API. GPU power and application throughput
metrics are queried from the Weights \& Biases API.

Conductor issues new commands through a distributed memory object cache that is
monitored by the management process on each node. Whenever the cache is updated
with a new power cap, the cap is enforced by launching \verb|nvidia-smi -pl| to
set a power limit on the compute node's GPUs. A MosaicML Composer callback
handles early stop requests received in the cache to allow jobs to checkpoint
immediately before suspension/rescaling.

\section{Conclusion}
This Phoenix-based field demonstration marks a transformative moment in
energy-AI integration. With solely a software-based solution, existing AI
clusters can become reliable, precise assets for grid support. As demand from
AI accelerates, flexible data centers offer a scalable, sustainable way to meet
electricity system needs--without delaying innovation or sacrificing
reliability.

The path forward will require deeper engagement with grid operators, AI
developers, and regulators to standardize these capabilities and unlock a new
era of grid-interactive computing.

\backmatter

\bmhead{Acknowledgements}
We would like to thank the following individuals for their insights and support throughout this work: Pradeep Vincent, Jay Jackson, Mike Sweeney, Andres Springborn, Brandon Records, Scott Campbell, Ron Caputo, Sara Chen, and Marcin Zablocki at Oracle; Jonathan Frankle, Yu-Lin Yang, Lauren Ladrech, and Stewart Sherpa at Databricks; Shanker Trivedi, Vlad Troy, Marc Spieler at NVIDIA; Rob Toews, Jordan Jacobs, David Katz, and Daniel Mulet at Radical Ventures; Markus Specks at Aventurine Partners; David Rousseau at SRP; Greg Bernosky, Chris Lynn, and Bruce Brazis at Arizona Public Service; Arushi Sharma Frank at Luminary; Tyler Norris at Duke University; Isaac Brown at 38 North; Sean Kelly at Amperon; Anna Patterson at Ceramic AI; Astrid Atkinson at Camus Energy; Gaurav Desai; Sherri Goodman; Max Boomer; Richard Stuebi at Boston University; and David Porter, Dave Larson, and Tom Wilson at EPRI.

\onecolumn\bibliography{bibliography}

\begin{thebibliography}{10}
\expandafter\ifx\csname url\endcsname\relax
  \def\url#1{\burl{#1}}\fi
\expandafter\ifx\csname urlprefix\endcsname\relax\def\urlprefix{URL }\fi
\providecommand{\bibinfo}[2]{#2}
\providecommand{\eprint}[2][]{\url{#2}}
\providecommand{\doi}[1]{\url{https://doi.org/#1}}
\bibcommenthead

\bibitem{johnson2025implications}
\bibinfo{author}{{National Academies of Sciences, Engineering, and Medicine}}.
\newblock \emph{\bibinfo{title}{Implications of Artificial
  Intelligence–Related Data Center Electricity Use and Emissions: Proceedings
  of a Workshop}}  (\bibinfo{publisher}{The National Academies Press},
  \bibinfo{address}{Washington, DC}, \bibinfo{year}{2025}).
\newblock \doi{10.17226/29101}.

\bibitem{bianchini2024data}
\bibinfo{author}{Ricardo Bianchini}, \bibinfo{author}{Christian Belady} \&
  \bibinfo{author}{Anand Sivasubramaniam}.
\newblock \bibinfo{title}{Data center power and energy management: Past,
  present, and future}.
\newblock \emph{\bibinfo{journal}{IEEE Micro}} \textbf{\bibinfo{volume}{44}},
  \bibinfo{pages}{30–36} (\bibinfo{year}{2024}).
\newblock \doi{10.1109/MM.2024.3426478}.

\bibitem{iea2025electricity}
\bibinfo{author}{Eren {\c{C}}am}, \bibinfo{author}{Marc Casanovas} \&
  \bibinfo{author}{John Moloney}.
\newblock \bibinfo{title}{Electricity 2025: Analysis and forecast to 2027}.
\newblock \bibinfo{howpublished}{IEA} (\bibinfo{year}{2025}).
\newblock \bibinfo{note}{\url{https://www.iea.org/reports/electricity-2025}}.

\bibitem{aljbour2024poweringintelligence}
\bibinfo{author}{Jordan Aljbour}, \bibinfo{author}{Tom Wilson} \&
  \bibinfo{author}{Poorvi Patel}.
\newblock \bibinfo{title}{Powering intelligence: Analyzing artificial
  intelligence and data center energy consumption}.
\newblock \bibinfo{howpublished}{Electric Power Research Institute ({EPRI})}
  (\bibinfo{year}{2024}).
\newblock
  \bibinfo{note}{\url{https://www.epri.com/research/products/000000003002028905}}.

\bibitem{zhang2022aqa}
\bibinfo{author}{Yijia Zhang}, \bibinfo{author}{Daniel~Curtis Wilson},
  \bibinfo{author}{Ioannis~Ch. Paschalidis} \& \bibinfo{author}{Ayse~K.
  Coskun}.
\newblock \bibinfo{title}{{HPC} data center participation in demand response:
  An adaptive policy with {QoS} assurance}.
\newblock \emph{\bibinfo{journal}{IEEE Transactions on Sustainable Computing}}
  \textbf{\bibinfo{volume}{7}}, \bibinfo{pages}{157--171}
  (\bibinfo{year}{2022}).
\newblock \doi{10.1109/TSUSC.2021.3077254}.

\bibitem{zhang2021realworld}
\bibinfo{author}{Yijia Zhang}, \bibinfo{author}{Daniel~C. Wilson},
  \bibinfo{author}{Ioannis~Ch. Paschalidis} \& \bibinfo{author}{Ayse~K. Coskun}
  \emph{\bibinfo{title}{A data center demand response policy for real-world
  workload scenarios in {HPC}}}.
\newblock In \emph{\bibinfo{booktitle}{2021 Design, Automation \& Test in
  Europe Conference \& Exhibition ({DATE})}}, \bibinfo{pages}{282--287}
  (\bibinfo{year}{2021}).

\bibitem{xing2023carbonresponder}
\bibinfo{author}{Jiali Xing}, \bibinfo{author}{Bilge Acun},
  \bibinfo{author}{Aditya Sundarrajan}, \bibinfo{author}{David Brooks},
  \bibinfo{author}{Manoj Chakkaravarthy}, \bibinfo{author}{Nikky Avila},
  \bibinfo{author}{Carole-Jean Wu} \& \bibinfo{author}{Benjamin~C. Lee}.
\newblock \bibinfo{title}{Carbon responder: Coordinating demand response for
  the datacenter fleet} (\bibinfo{year}{2023}).
\newblock
  \bibinfo{eprint}{{\href{https://arxiv.org/abs/2311.08589}{{arXiv:2311.08589}}}}.

\bibitem{radovanovic2023carbon}
\bibinfo{author}{Ana Radovanović}, \bibinfo{author}{Ross Koningstein},
  \bibinfo{author}{Ian Schneider}, \bibinfo{author}{Bokan Chen},
  \bibinfo{author}{Alexandre Duarte}, \bibinfo{author}{Binz Roy},
  \bibinfo{author}{Diyue Xiao}, \bibinfo{author}{Maya Haridasan},
  \bibinfo{author}{Patrick Hung}, \bibinfo{author}{Nick Care},
  \bibinfo{author}{Saurav Talukdar}, \bibinfo{author}{Eric Mullen},
  \bibinfo{author}{Kendal Smith}, \bibinfo{author}{MariEllen Cottman} \&
  \bibinfo{author}{Walfredo Cirne}.
\newblock \bibinfo{title}{Carbon-aware computing for datacenters}.
\newblock \emph{\bibinfo{journal}{IEEE Transactions on Power Systems}}
  \textbf{\bibinfo{volume}{38}}, \bibinfo{pages}{1270--1280}
  (\bibinfo{year}{2023}).
\newblock \doi{10.1109/TPWRS.2022.3173250}.

\bibitem{sivaram2018taming}
\bibinfo{author}{Varun Sivaram}.
\newblock \emph{\bibinfo{title}{Taming the Sun: Innovations to Harness Solar
  Energy and Power the Planet}}  (\bibinfo{publisher}{The MIT Press},
  \bibinfo{address}{Cambridge, Massachusetts}, \bibinfo{year}{2018}).
\newblock \bibinfo{note}{ISBN: 978-0-262-53707-0}.

\bibitem{tansoo2024usingcba}
\bibinfo{author}{Jie-Sheng {Tan-Soo}}, \bibinfo{author}{Ping {Qin}},
  \bibinfo{author}{Yifei {Quan}}, \bibinfo{author}{Jun {Li}} \&
  \bibinfo{author}{Xiaoxi {Wang}}.
\newblock \bibinfo{title}{Using cost{\textendash}benefit analyses to identify
  key opportunities in demand-side mitigation}.
\newblock \emph{\bibinfo{journal}{Nature Climate Change}}
  \textbf{\bibinfo{volume}{14}}, \bibinfo{pages}{1158--1164}
  (\bibinfo{year}{2024}).
\newblock \doi{10.1038/s41558-024-02146-4}.

\bibitem{norris2025rethinking}
\bibinfo{author}{Tyler Norris}, \bibinfo{author}{Timothy Profeta},
  \bibinfo{author}{Dalia Patino-Echeverri} \& \bibinfo{author}{Adam
  Cowie-Haskell}.
\newblock \bibinfo{title}{Rethinking load growth: assessing the potential for
  integration of large flexible loads in {US} power systems}.
\newblock \bibinfo{howpublished}{Nicholas Institute for Energy, Environment \&
  Sustainability} (\bibinfo{year}{2025}).
\newblock \bibinfo{note}{\url{https://hdl.handle.net/10161/32077}}.

\bibitem{caiso2021rca}
\bibinfo{author}{Elliot Mainzer}, \bibinfo{author}{Marybel Batjer} \&
  \bibinfo{author}{David Hochschild}.
\newblock \bibinfo{title}{Final root cause analysis: Mid-august 2020 extreme
  heat wave}.
\newblock \bibinfo{howpublished}{California Independent System Operator
  ({CAISO})} (\bibinfo{year}{2021}).
\newblock
  \bibinfo{note}{\url{https://www.caiso.com/Documents/Final-Root-Cause-Analysis-Mid-August-2020-Extreme-Heat-Wave.pdf}}.

\bibitem{lin2024explodingai}
\bibinfo{author}{Liuzixuan Lin}, \bibinfo{author}{Rajini Wijayawardana},
  \bibinfo{author}{Varsha Rao}, \bibinfo{author}{Hai Nguyen},
  \bibinfo{author}{Emmanuel~Wedan GNIBGA} \& \bibinfo{author}{Andrew~A. Chien}
  \emph{\bibinfo{title}{Exploding {AI} power use: an opportunity to rethink
  grid planning and management}}.
\newblock In \emph{\bibinfo{booktitle}{Proceedings of the 15th ACM
  International Conference on Future and Sustainable Energy Systems}}, e-Energy
  '24, \bibinfo{pages}{434–441} (\bibinfo{publisher}{Association for
  Computing Machinery}, \bibinfo{address}{New York, NY, USA},
  \bibinfo{year}{2024}).
\newblock \urlprefix\url{https://doi.org/10.1145/3632775.3661959}.

\bibitem{zhao2023sustainable}
\bibinfo{author}{Dan Zhao}, \bibinfo{author}{Siddharth Samsi},
  \bibinfo{author}{Joseph McDonald}, \bibinfo{author}{Baolin Li},
  \bibinfo{author}{David Bestor}, \bibinfo{author}{Michael Jones},
  \bibinfo{author}{Devesh Tiwari} \& \bibinfo{author}{Vijay Gadepally}
  \emph{\bibinfo{title}{Sustainable supercomputing for {AI}: {GPU} power
  capping at {HPC} scale}}.
\newblock In \emph{\bibinfo{booktitle}{Proceedings of the 2023 ACM Symposium on
  Cloud Computing}}, SoCC '23, \bibinfo{pages}{588–596}
  (\bibinfo{publisher}{Association for Computing Machinery},
  \bibinfo{address}{New York, NY, USA}, \bibinfo{year}{2023}).
\newblock \urlprefix\url{https://doi.org/10.1145/3620678.3624793}.

\bibitem{lechowicz2023pauseresume}
\bibinfo{author}{Adam Lechowicz}, \bibinfo{author}{Nicolas Christianson},
  \bibinfo{author}{Jinhang Zuo}, \bibinfo{author}{Noman Bashir},
  \bibinfo{author}{Mohammad Hajiesmaili}, \bibinfo{author}{Adam Wierman} \&
  \bibinfo{author}{Prashant Shenoy}.
\newblock \bibinfo{title}{The online pause and resume problem: Optimal
  algorithms and an application to carbon-aware load shifting}.
\newblock \emph{\bibinfo{journal}{Proc. ACM Meas. Anal. Comput. Syst.}}
  \textbf{\bibinfo{volume}{7}} (\bibinfo{year}{2023}).
\newblock \doi{10.1145/3626776}.

\bibitem{ratnayake2025gridflexibilityneeds}
\bibinfo{author}{Anuja Ratnayake}, \bibinfo{author}{Irene~Danti Lopez},
  \bibinfo{author}{Baskar Vairamohan} \& \bibinfo{author}{Eamonn Lannoye}.
\newblock \bibinfo{title}{Grid flexibility needs and data center
  characteristics}.
\newblock \bibinfo{howpublished}{Electric Power Research Institute ({EPRI})}
  (\bibinfo{year}{2025}).
\newblock
  \bibinfo{note}{\url{https://www.epri.com/research/programs/063638/results/3002031504}}.

\bibitem{satchwell2025electricity}
\bibinfo{author}{Andrew Satchwell}, \bibinfo{author}{Natalie~Mims Frick},
  \bibinfo{author}{Peter Cappers}, \bibinfo{author}{Sanem Sergici},
  \bibinfo{author}{Ryan Hledik}, \bibinfo{author}{Goksin Kavlak} \&
  \bibinfo{author}{Glenda Oskar}.
\newblock \bibinfo{title}{Electricity rate designs for large loads: Evolving
  practices and opportunities}.
\newblock \bibinfo{howpublished}{Lawrence Berkeley National Laboratory}
  (\bibinfo{year}{2025}).
\newblock \bibinfo{note}{Retrieved from
  \url{https://escholarship.org/uc/item/19m1k8vr}}.

\bibitem{zheng2020mitigatingcurtailment}
\bibinfo{author}{Jiajia Zheng}, \bibinfo{author}{Andrew~A. Chien} \&
  \bibinfo{author}{Sangwon Suh}.
\newblock \bibinfo{title}{Mitigating curtailment and carbon emissions through
  load migration between data centers}.
\newblock \emph{\bibinfo{journal}{Joule}} \textbf{\bibinfo{volume}{4}},
  \bibinfo{pages}{2208--2222} (\bibinfo{year}{2020}).
\newblock \doi{10.1016/j.joule.2020.08.001}.

\bibitem{mehra2023supporting}
\bibinfo{author}{Varun Mehra} \& \bibinfo{author}{Raiden Hasegawa}.
\newblock \bibinfo{title}{Supporting power grids with demand response at google
  data centers}.
\newblock \bibinfo{howpublished}{Google Cloud Blog} (\bibinfo{year}{2023}).
\newblock
  \bibinfo{note}{\url{https://cloud.google.com/blog/products/infrastructure/using-demand-response-to-reduce-data-center-power-consumption}}.

\bibitem{zhou2016survey}
\bibinfo{author}{Zhi Zhou}, \bibinfo{author}{Todd Levin} \&
  \bibinfo{author}{Guenter Conzelmann}.
\newblock \bibinfo{title}{Survey of {U.S.} ancillary services markets}.
\newblock \bibinfo{howpublished}{Argonne National Lab ({ANL})}
  (\bibinfo{year}{2016}).
\newblock \bibinfo{note}{\url{https://www.osti.gov/biblio/1236451}}.

\bibitem{ercot2024ancillary}
\bibinfo{title}{{ERCOT} ancillary services study} (\bibinfo{year}{2024}).
\newblock \bibinfo{note}{Whitepaper retrieved from
  \url{https://www.ercot.com/files/docs/2024/10/07/ERCOT-Ancillary-Services-Study-Final-White-Paper.pdf}}.

\end{thebibliography}
%%\input article.bbl

\end{document}